\documentclass[final,5p,times,twocolumn]{elsarticle}

\usepackage{amssymb}
\usepackage{dblfloatfix}

\journal{Physics Letters B}

\begin{document}

\begin{frontmatter}



%\title{}
\title{Assessing the impact of valence {\em sd} neutrons and protons on fusion}


 \author{Varinderjit Singh}
 \author{J. Vadas}
 \author{T.K. Steinbach}%
 \author{B.~B. Wiggins}
 \author{S. Hudan}
\author{R.~T. deSouza\corref{cor1}\fnref{label1}}
%\author{R.~T. deSouza}
\cortext[cor1]{E-mail: deSouza@indiana.edu}
\fntext[label1]{Corresponding author}
%\ead{desouza@indiana.edu}
\address{Department of Chemistry and Center for Exploration of Energy and Matter, Indiana University\\
2401 Milo B. Sampson Lane, Bloomington, IN 47408, USA}

\begin{abstract}
  Experimental near-barrier fusion cross-sections for $^{17}$F + $^{12}$C
  are compared to the fusion excitation
functions for $^{16,18}$O, $^{19}$F, and $^{20}$Ne ions on a carbon target. 
Comparison of the reduced fusion cross-section for
the different systems accounts for the differing 
static size of the incident ions and changes in fusion barrier.
Remaining trends of the fusion cross-section above the
barrier are observed. These trends are interpreted as the 
interplay of the {\em sd} protons and neutrons.
The experimental data are also compared to
a widely-used analytic model of near-barrier fusion,
a time-dependent Hartree-Fock model, and coupled channels calculations.
\end{abstract}


\end{frontmatter}


\section{Introduction}
%\label{}

Nuclear fusion is a topic of considerable interest both from a fundamental perspective as well as in the field of nuclear astrophysics \cite{Back14}. Nuclei 
just beyond a closed shell present a unique opportunity to probe the interplay of shell and collective effects on the fusion process. In particular, 
light nuclei just beyond the 1p$_{1/2}$ shell,  
namely isotopes of oxygen, fluorine, and neon are 
good candidates for examination. In this work, the fusion of various isotopes of these 
elements with a carbon target at near-barrier energies is examined. The results of this work, 
which combines both stable and radioactive beams, points to the potential 
of low-energy beams at radioactive beam facilities \cite{GANIL, FRIB}
for examining the impact of neutron-excess on fusion. 

Addition of neutrons and protons just beyond the 1p$_{1/2}$ shells of $^{16}$O clearly changes both the matter and charge distributions of the nuclei. Theoretical calculations indicate that for a large neutron excess, 
e.g. $^{24}$O as compared to $^{16}$O, fusion with $^{16}$O target is significantly enhanced \cite{Umar12}. The impact of adding just a few neutrons or protons 
beyond the 1p$_{\frac{1}{2}}$ shell on fusion is less clear.
With increased 
atomic or mass number, 
the fusion barrier and consequently the fusion cross-section is clearly impacted. We propose to go beyond these trivial systematic differences and examine
the detailed differences in the fusion cross-section.
A particularly interesting case to investigate is fusion of
the nucleus $^{17}$F 
which exhibits a proton-halo when in its 2s$_{1/2}$ excited state
\cite{Morlock97}. It is presently unclear whether an increased radial
extent results in a fusion enhancement or weak binding results in a decreased fusion cross-section. For the case of
$^{17}$F + $^{208}$Pb, neither an enhancement nor a suppression of
fusion was observed relative to $^{19}$F + $^{208}$Pb
\cite{Rehm98}. However, in the case of fusion with a large target nucleus,
such as $^{208}$Pb, the
impact of adding two neutrons on fusion might be diminished.
To examine the impact on fusion of adding a few protons and neutrons to the
{\em sd} shell we measure fusion for $^{17}$F + $^{12}$C and compare it with fusion
induced by O, F, and Ne beams on $^{12}$C.

\section{Experimental Details}

Discovery of halo nuclei, notably $^{11}$Li~\cite{Tanihata85a, Tanihata85b}
was achieved through systematic examination of the interaction cross-sections for 
lithium isotopes. 
At high incident energy one
expects the sudden approximation to be valid in describing the collision of the nuclei.
Consequently, the nuclear densities do not have enough time to rearrange 
as the projectile and target nuclei come into contact. Thus a measurement of the interaction
cross-section at high energy probes the nuclear size and other geometrical features such as 
deformation, all of which are considered ``static'' \cite{Tanihata85a}. 
Hence, the measured interaction 
cross-section, $\sigma_{I}$, 
provides a direct and effective measure 
of the extent of the matter distribution.

To better understand the change in the static size of the different nuclei
considered in this work,
we examine the interaction cross-sections measured at high energy.
Presented in Table~\ref{tab:IntXS} are $\sigma_I$  
for O, F, and Ne nuclei with a carbon target \cite{Ozawa01a} with
the number of protons and neutrons in the {\em sd} shell indicated. 
The closure of the 1p$_{\frac{1}{2}}$ with N=8 provides a natural reference 
from which to examine the impact made by the presence of a few nucleons in the {\em sd} shell. 
Addition of two neutrons to $^{16}$O, $^{17}$F, and $^{18}$Ne increases 
the interaction cross-section by 50, 61, and 68 mb respectively. Addition of 
two protons to $^{16}$O, i.e. $^{18}$Ne, results in an increase of
$\sigma_{I}$ by 94 mb while in the case $^{18}$O to $^{20}$Ne, it increases by 112 mb.
The difference in the interaction cross-section with  the addition of two protons (94-112 mb)
as compared to addition of two neutrons (50-68 mb) is presumably due to the repulsion
of the two protons with the $^{16}$O core. Such a difference is not captured by systematics which relate the radius
of a nucleus to simply the mass number, A, suggesting the advantage of using
$\sigma_I$ to account for trivial size effects.

\begin{table}
  \center
   \centering
   \caption{Interaction cross-sections for oxygen, fluorine, and neon nuclei with a carbon target
     at an energy about 900 A MeV. Taken from \cite{Ozawa01a}.}
   \label{tab:IntXS}
   \begin{tabular}{|c|c|}
     \hline
     $^{18}$Ne$^{2p0n}$   & $^{20}$Ne$^{2p2n}$\\
     
     $\sigma_I$=1076$\pm$25 mb & $\sigma_I$=1144$\pm$10 mb\\
     \hline
     $^{17}$F$^{1p0n}$     & $^{19}$F$^{1p2n}$\\
     $\sigma_I$=982$\pm$32 mb & $\sigma_I$=1043$\pm$24 mb\\
     \hline
     $^{16}$O$^{0p0n}$     & $^{18}$O$^{0p2n}$\\
     $\sigma_I$=982$\pm$6 mb & $\sigma_I$=1032$\pm$26 mb\\
     \hline
     
\end{tabular}
\end{table}

While the excitation function for fusion of $^{16,18}$O, $^{19}$F, and $^{20}$Ne ions with a carbon target 
already exists, no data exists for $^{17}$F+$^{12}$C
or $^{18}$Ne+$^{12}$C. 
A beam of $^{16}$O ions at Florida State University was used to  
produce $^{17}$F ions via 
a (d,n) reaction. These ions were separated from the incident beam by the electromagnetic spectrometer 
RESOLUT \cite{RESOLUT}. 
The beam exiting the spectrometer 
consisted of both $^{17}$F and residual $^{16}$O ions necessitating identification of each ion on a 
particle-by-particle basis. The presence of both species allowed the 
simultaneous 
measurement of $^{16}$O~+~$^{12}$C 
and $^{17}$F~+~$^{12}$C providing a built-in reference measurement and 
comparison with the 
well-established fusion excitation function for
$^{16}$O~+~$^{12}$C.

\begin{figure}[]
\includegraphics[scale=0.7]{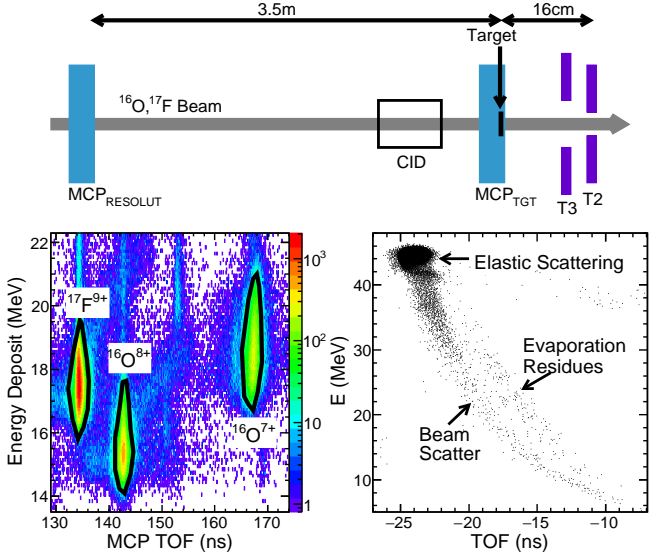}
\caption{Schematic of the experimental setup along with a PID and ETOF spectra. Particle identification of the $^{17}$F and $^{16}$O ions incident on the target is displayed in the left plot along with a representative ETOF spectrum (selected on $^{17}$F) used to identify evaporation residues (ER) in the right plot. 
}
\label{fig:setup}
\end{figure}

\begin{figure}[t]
\includegraphics[scale=0.40]{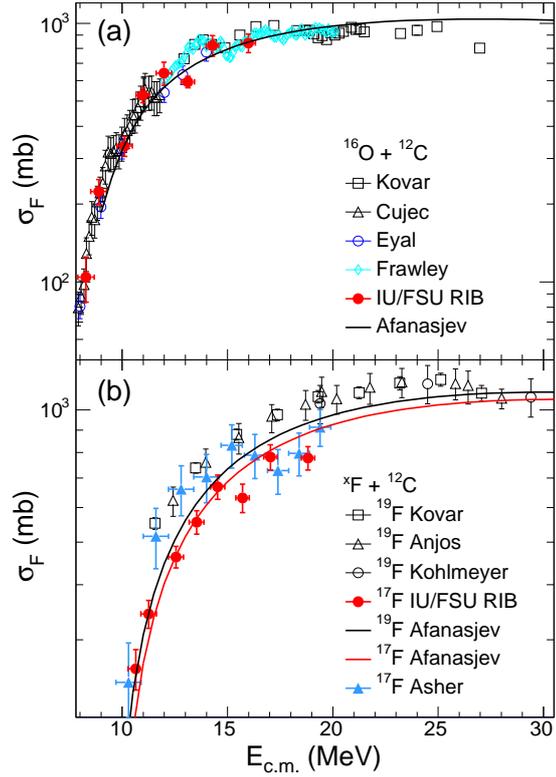}
 \caption{Fusion excitation functions for $^{16}$O + $^{12}$C (top panel) and $^{17,19}$F + $^{12}$C (bottom panel). The predictions of an analytic fusion model \cite{Afanasjev12} are indicated by the 
   solid lines.
 }
 \label{fig:O16Fxsect}
 \end{figure}

The setup used to measure fusion of fluorine and oxygen ions with carbon 
nuclei in this experiment 
is depicted in Fig.~\ref{fig:setup}. Two  microchannel plate detectors
designated MCP$\mathrm{_{RESOLUT}}$ and MCP$\mathrm{_{TGT}}$ spaced $\sim$3.5 m 
apart and a compact 
ionization detector (CID) permitted a $\Delta$E-TOF measurement for each ion incident on the target. 
Measurement of the 
$\Delta$E-TOF (Fig.~\ref{fig:setup}) clearly shows
three peaks corresponding to $^{17}$F$^{9+}$, $^{16}$O$^{7+}$, and $^{16}$O$^{8+}$ ions. The intensity of the $^{17}$F beam incident on the target was 
3-7 x 10$^{3}$ ions/s with a purity of 37\%-54\%.
 
Fusion of a $^{17}$F ($^{16}$O) nucleus in the beam together with a $^{12}$C nucleus in the target 
foil results in the production of an excited $^{29}$P 
($^{28}$Si) nucleus. For near-barrier collisions the excitation of the fusion product is relatively modest, E$^*$ $\approx$ 30 MeV. 
Emission of particles as the fusion 
product de-excites, deflects the evaporation residue (ER) from 
the beam direction. This deflection 
allows its detection and identification using two annular silicon detectors 
designated T2 and T3 that subtend the angular range 
3.5$^\circ$ $<$ $\theta_{lab}$ $<$ 25$^\circ$. 
Using the measured energy deposit in the silicon detectors  
and the time-of-flight \cite{deSouza11}, the mass of the ion was calculated. ERs 
were cleanly distinguished from beam particles by their mass\cite{Steinbach14,Steinbach14a}. A representative ETOF spectrum is presented in Fig.~\ref{fig:setup}.

The fusion cross-section, $\sigma_F$ is extracted from the measured yield of evaporation residues through the 
relation $\sigma_F$ = N$_{ER}$/($\epsilon_{ER}$ x t x N$_{I}$) where 
N$_{I}$ is the number of beam particles of a given type incident on the target, t is the target 
thickness, $\epsilon_{ER}$ is the detection efficiency, and N$_{ER}$ 
is the number of evaporation residues detected. The number N$_{I}$ is determined by counting the 
particles with the appropriate time-of-flight between the two microchannel plates 
that additionally have the correct identification in the $\Delta$E-TOF map 
depicted in Fig.~\ref{fig:setup}. 
The target thickness, t, is 105 $\pm$ 0.5 $\mu$g/cm$^2$. The number of detected residues, 
N$_{ER}$, is determined by summing the number of detected residues clearly identified by the ETOF technique as shown in Fig.~\ref{fig:setup} 
\cite{Steinbach14a}. 
To obtain the detection efficiency, $\epsilon_{ER}$,
a statistical model is 
used to describe the de-excitation of the fusion product together with the geometric acceptance of the experimental setup. 
The detection efficiency varied from $\sim$81\% at the highest incident energies measured 
to $\sim$85\% at the 
lowest incident energy due to the changing kinematics of the reaction with an associated uncertainty of approximately 5$\%$.

\begin{figure}[t]
\includegraphics[scale=0.40]{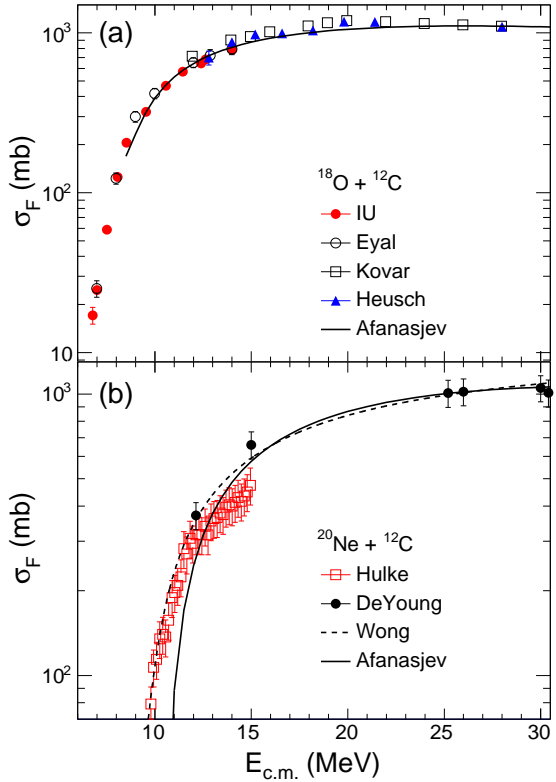}
\caption{Fusion excitation functions for $^{18}$O+$^{12}$C (top panel) and $^{20}$Ne+$^{12}$C (bottom panel) together with the predictions of the analytic model\cite{Afanasjev12}.
The fit of the $^{20}$Ne data by a one-dimensional barrier-penetration model \cite{Wong73} is depicted by the dashed line.
}
\label{fig:O18Nexsect}
\end{figure}

\section{Results and Discussion}

Depicted in Fig.~\ref{fig:O16Fxsect} 
are the measured excitation functions for
fusion of $^{16}$O and $^{17}$F ions with a carbon target. With the experimental technique used, both
the $^{16}$O + $^{12}$C and $^{17}$F + $^{12}$C excitation functions are simultaneously measured. 
In the top panel, for $^{16}$O, one observes the good agreement 
of this measurement, depicted as the solid (red) circles, with the experimental 
cross-sections reported in the literature \cite{Cujec76, Eyal76, Kovar79, Frawley82}. The present measurement reproduces the known resonances in the interval 
10 MeV $\le$E$_{c.m.}$$\le$ 18 MeV demonstrating that the measurement of the fusion excitation function for $^{17}$F~+~$^{12}$C is robust.
In the lower panel of Fig.~\ref{fig:O16Fxsect} the measured cross-sections 
for $^{17}$F+$^{12}$C (solid red circles) are presented along with the excitation function 
for $^{19}$F+$^{12}$C. The multiple measurements for $^{19}$F \cite{Kohlmeyer77, Kovar79, Anjos90} 
are in agreement within the measurement 
uncertainties. As might be expected naively by the reduction of two neutrons, the $^{17}$F-induced fusion of this measurement exhibits a lower cross-section
than $^{19}$F for all energies shown.
The measured excitation functions are compared with the predictions of an analytic model based on a parameterization of the Sao Paulo potential
model coupled with a barrier penetration formalism \cite{Afanasjev12, Beard10, Yakolev10}. This model which has parameterized a large number of reactions
for low and mid-mass systems is a useful tool for network simulations in the near-barrier regime.
Despite having several fitted parameters, the analytic model has no
adjustable parameters and thus serves as a useful benchmark of the
expected systematic behavior.
The above-barrier cross-sections predicted by this model are depicted in Fig.~\ref{fig:O16Fxsect} as the solid lines. 
In the case of the $^{16}$O fusion the analytic model provides a reasonable 
description of the overall behavior of the excitation function. The failure of the analytic 
model to reproduce the resonance is unsurprising as it assumes a smooth, structureless barrier.
For the fluorine isotopes, while the excitation function for $^{17}$F is reasonably described, 
the excitation function for $^{19}$F is clearly underpredicted in the energy range measured. Nonetheless, the analytic
model does predict an increase in the cross-section associated with the presence of the two additional
neutrons in $^{19}$F as compared to $^{17}$F.

Also presented in Fig.~\ref{fig:O16Fxsect} is the excitation function for $^{17}$F + $^{12}$C recently measured by
Asher et al. (solid blue triangles) \cite{Asher20}. The statistical uncertainties of the Asher measurement are significantly
larger than those of the  present measurement. As importantly, the Asher measurement has
significantly larger uncertainty in the energy, an inherent consequence of the technique utilized. Even
considering the uncertainties of both excitation functions it is clear that the Asher excitation function
reports larger cross-sections or is shifted to lower energies as compared to the present measurement. Surprisingly,
the excitation function measured by Asher for E$_{c.m.}$$\le$15 MeV
is essentially consistent
with the excitation function for $^{19}$F. This result would imply that the removal of two neutrons from
$^{19}$F makes a negligible difference in both the barrier and the size of the fusing system. Such a result is inconsistent with the analytic model results which describes the expected systematic behavior. Consequently, for the remainder of this work only the present
$^{17}$F data is considered.

The excitation functions for fusion of $^{18}$O and $^{20}$Ne nuclei with 
carbon are shown 
in Fig.~\ref{fig:O18Nexsect}. 
While good agreement is observed for the various $^{18}$O datasets, 
in the case of $^{20}$Ne, for 12 MeV 
$\le$E$_{c.m.}$$\le$15 MeV 
a discrepancy exists between the measurements of 
Hulke \cite{Hulke80}  and deYoung \cite{deYoung82}. This discrepancy at E$_{c.m.}$ $\sim$15 MeV is larger than the reported uncertainties by both experimental measurements. Moreover, the shape of the excitation function determined by Hulke~\cite{Hulke80} deviates from the behavior of a smooth barrier as indicated by the 
analytic model \cite{Afanasjev12}. The Hulke data~\cite{Hulke80} for 
E$_{c.m.}$$\le$11 MeV and the deYoung data \cite{deYoung82} can be described by a
one-dimensional barrier penetration model \cite{Wong73} as indicated by the dashed line in 
Fig.~\ref{fig:O18Nexsect}.

\begin{figure}[]
\includegraphics[scale=0.7]{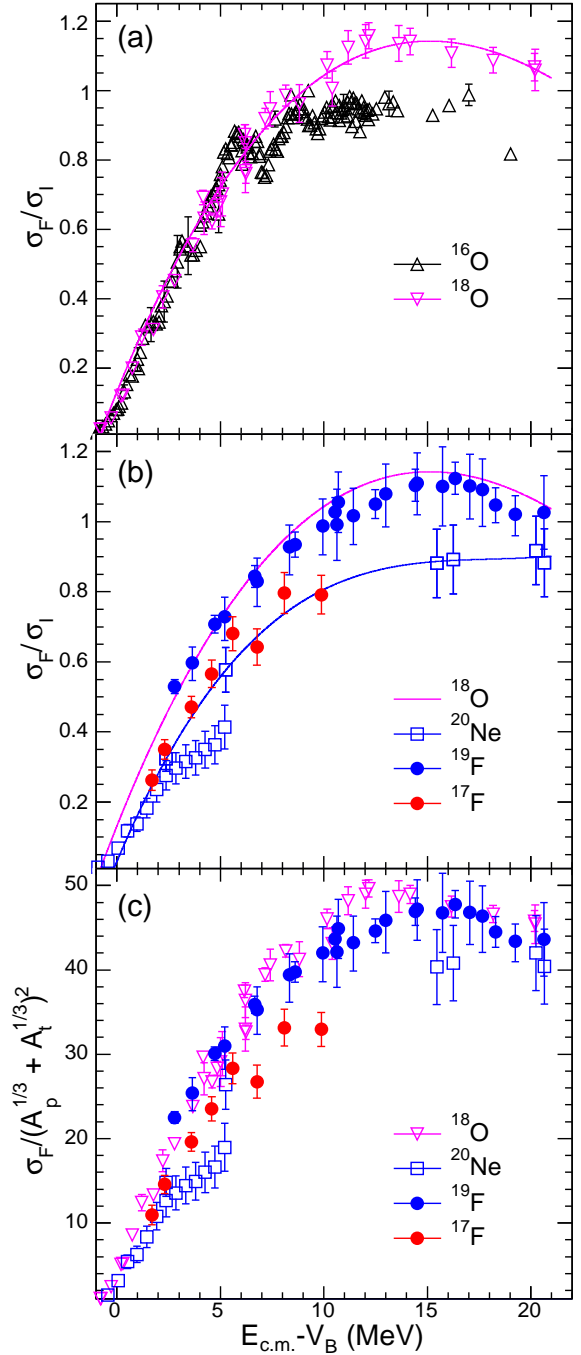}
\caption{Comparison of the reduced excitation functions for fusion of $^{16}$O and $^{18}$O 
ions (top panel) and $^{18}$O, $^{20}$Ne, $^{19}$F, and $^{17}$F ions (lower panels) on a carbon target. The line through the $^{20}$Ne data simply serves to guide the eye.   
}
\label{fig:Red_xsect}
\end{figure}

In order to appropriately compare all the fusion excitation functions for these light nuclei, we scale the
fusion cross-section by the interaction cross-section, $\sigma_I$, presented in Table ~\ref{tab:IntXS}. 
In addition, the trivial effect of the 
different barriers is accounted for by examining the reduced cross-section, $\sigma_F$/$\sigma_I$, as
a function of the above-barrier energy, E$_{c.m.}$-V$_B$. 
The value of the 
barrier, V$_B$,  is taken from the Bass model \cite{Bass80}. The 
uncertainties shown in the reduced
cross-section reflect the uncertainties in both $\sigma_F$ and $\sigma_I$.
This presentation
allows one to investigate differences between the nuclei shown after 
effectively eliminating systematic differences in the static size and barrier.
Comparison of all the excitation functions in Fig.~\ref{fig:Red_xsect} yields 
some interesting results. In Fig.~\ref{fig:Red_xsect}a 
one observes that for 
(E$_{c.m.}$-V$_B$)$\le$9 MeV, 
the reduced cross-sections for $^{16}$O and $^{18}$O are relatively comparable. 
The prominent resonance structure for $^{16}$O-induced fusion 
is absent in the 
case of $^{18}$O. Above 9 MeV one observes that the reduced cross-section for $^{18}$O exceeds that of $^{16}$O. This increase 
in the reduced cross-section indicates the impact of the two 
{\em sd} neutrons on the fusion cross-section 
{\em over and above the increase in the static size}. In further understanding the role 
of { \em sd} valence nucleons on fusion, we elect to choose $^{18}$O as our reference. 
The absence of strong, sharp resonance structures in the $^{18}$O reaction supports this choice. To facilitate the use of $^{18}$O as a reference we have described the data  
by the smooth curve depicted in Fig.~\ref{fig:Red_xsect}a. This curve 
corresponds to a third-order polynomial fit to the reduced cross-sections 
for $^{18}$O and simply serves as an adequate representation of the 
$^{18}$O data in Fig.~\ref{fig:Red_xsect}b.

Presented in Fig.~\ref{fig:Red_xsect}b are the reduced excitation functions for 
$^{19}$F, $^{20}$Ne, and $^{17}$F in comparison to $^{18}$O. Within the measurement uncertainties the 
$^{19}$F data manifests the same reduced cross-section as the $^{18}$O. However, there is a subtle
indication that the reduced cross-section for $^{19}$F appears to be systematically slightly lower
than that of $^{18}$O 
for (E$_{c.m.}$-V$_B$)$\ge$9 MeV. The $\langle$$\sigma_F$/$\sigma_I$$\rangle$
in this energy interval is 1.11$\pm$0.04 for $^{18}$O and  1.06$\pm$0.07 for $^{19}$F. This
comparison between
$^{18}$O and $^{19}$F is complicated by the presence of the unpaired proton in $^{19}$F.

Examination of 
the reduced cross-section for $^{20}$Ne 
is more telling. For all energies measured, the $^{20}$Ne reduced cross-section is lower than that of both
$^{18}$O and $^{19}$F. 
Unfortunately, lack of data between $\sim$5 MeV$\le$(E$_{c.m.}$-V$_B$)$<$15 MeV 
prevents a better characterization of this excitation function.
The reduced cross-sections for $^{17}$F observed in this work are reduced as
compared to the cross-sections for $^{19}$F for the entire energy range
measured. 
This decreased value for $^{17}$F as compared to 
$^{19}$F is qualitatively 
consistent with the trend observed for $^{18}$O and $^{16}$O.

To investigate the interaction of the valence protons on the valence 
neutrons we have performed
Relativistic Mean Field (RMF) calculations of the neutron and proton density 
distributions of light nuclei.
In examining the proton and neutron density distributions 
for neutron-rich carbon nuclei from these RMF calculations, one observes that 
with increasing neutron number not only does the tail of the neutron density 
distribution extend further out 
but despite a constant number of protons, the proton density distribution 
is slightly 
extended \cite{deSouza20}. This interaction between valence protons  
and neutrons is also reflected in the one proton separation energies of 
$^{16}$O,$^{18}$O,and $^{20}$O which are 12.1, 15.9, and 19.3 MeV respectively.

Based upon these calculations we hypothesize that the  slight decrease in the
cross-section for $^{19}$F as compared to $^{18}$O could
reflect the 
attraction of the {\em sd} proton on the two {\em sd} neutrons resulting in a
reduction of the fusion probability.
The lower cross-sections for $^{20}$Ne as compared to $^{18}$O and $^{19}$F
is similarly consistent with the 
attraction of the two {\em sd} protons on the two {\em sd} neutrons and the consequent suppression of fusion. It should be appreciated that the explanation of these observations is at a qualitative level and a more quantitative description
will require more sophisticated theoretical calculations which include nuclear
structure. 

These observations for the fusion of light {\em sd} nuclei 
can be summarized as follows.
The fusion cross-section is enhanced by the 
presence of {\em sd} neutrons. This increased cross-section is suppressed  
by the presence of {\em sd} protons. It should be stressed that these changes
in the reduced fusion cross-section relative to the barrier are beyond the 
systematic changes expected.
It is clear that further experimental data, particularly in the case of $^{20}$Ne and $^{17}$F 
would be extremely useful. 
To assess the sensitivity of our results to the use of the interaction 
cross-section, $\sigma_I$, as a reference, we have calculated the 
reduced cross-section as 
$\sigma_F$/(A$_P$$^{\frac{1}{3}}$+A$_T$$^{\frac{1}{3}}$)$^2$ where 
A$_P$ and A$_T$ are the mass number of the projectile and target nuclei respectively.
While more sophisticated prescriptions for the reduced cross-section have been utilized in
comparing dissimilar systems, 
given the similarity of the systems compared in this work this simple scaling is appropriate \cite{Canto15}.
Examining the 
dependence of this quantity on (E$_{c.m.}$-V$_B$) in Fig.~\ref{fig:Red_xsect}c
reveals that although the 
reduced cross-section for $^{20}$Ne lies closer to that of $^{19}$F, the 
trends observed in Fig.~\ref{fig:Red_xsect}b
remain the same indicating the robustness of the observations.

\begin{figure}[t]
\includegraphics[scale=0.40]{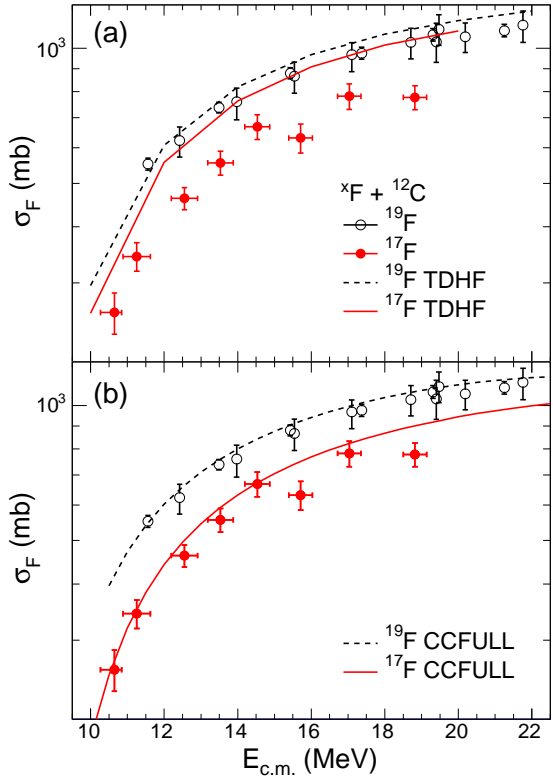}
\caption{Comparison of the fusion excitation function for 
  $^{17,19}$F+$^{12}$C with the predictions of a TDHF model (panel a) and
  CCFULL (panel b). 
  }
\label{fig:TDHF_xsect}
\end{figure}

To investigate the role of dynamics in the $^{17,19}$F + $^{12}$C reaction we performed 
time-dependent Hartree Fock calculations.
On general grounds the TDHF
approach is well-suited to describing the 
large-amplitude collective motion associated with fusion. 
Artificial symmetry restrictions are eliminated by performing the TDHF
calculations on a 3D cartesian grid~\cite{Umar06a}. Recent 
calculations~\cite{Kim97a} have provided a good description of 
above-barrier fusion data~\cite{Simenel08,Simenel17}.
The TDHF calculations were performed by initiating 
collisions for increasing impact parameters in steps of 0.01~fm
until the maximum impact parameter for fusion is reached.
Calculations were
performed using the Sky3D model~\cite{Sky3D}, with a SV-bas 
interaction with DDDI pairing. It should be noted that pairing significantly impacts the
fusion cross-section in TDHF \cite{Steinbach14a} emphasizing the need for its accurate description.
The results of the TDHF calculations for $^{17}$F and $^{19}$F are depicted in 
Fig.~\ref{fig:TDHF_xsect}a
as the solid and dashed lines respectively. While the model provides a reasonable description of the $^{19}$F data, the $^{17}$F cross-sections are overpredicted. It should be noted that TDHF 
calculations often slightly overpredict the above-barrier cross-sections due to their neglect of breakup processes, complicating direct comparison of the cross-section.
Nonetheless, it is 
noteworthy that the difference between $^{19}$F and $^{17}$F observed experimentally is not reproduced. 

Coupled-channels calculations \cite{Dasso83, Esbensen08} have been performed to assess the relative
importance of the ground and excited states on the fusion cross-section. Results of
CCFULL calculations \cite{Hagino99} in which only the ground state of both target and projectile
nuclei is considered are presented in
Fig.~\ref{fig:TDHF_xsect}b. A reasonable description of both
excitation functions is achieved. Inclusion of excited states acts to increase the cross-section slightly,
though primarily at sub-barrier energies. The limited impact of excited states on fusion of $^{17}$F in this near-barrier regime
is consistent with other measurements \cite{Asher20, Rehm98}.
This description of the measured cross-sections corresponds to a potential
in CCFULL with R$_B$= 7.89 fm, V$_B$=9.17 MeV, and $\hbar$$\omega$=3.36 MeV for $^{17}$F and
R$_B$= 8.42 fm, V$_B$=8.58 MeV, and $\hbar$$\omega$=2.94 MeV for $^{19}$F.
While the limited span of the experimental data does not permit a reliable
extraction of $\hbar$$\omega$, the trend of increasing R$_B$ and decreasing V$_B$
with the addition of two neutrons is qualitatively understandable reflecting the
increased nuclear interaction.

\section{Summary}
Systematic comparison of the fusion excitation functions for isotopes of O, F, and Ne 
nuclei with a carbon target can be used to examine the impact of valence {\em sd} protons and neutrons
on fusion. 
After accounting for differences in the static size of the incident nuclei
and systematic changes in the fusion barrier, 
differences are noted between fusion of $^{18}$O, $^{19}$F, $^{17}$F, and $^{20}$Ne with a carbon target.
At energies about 5 MeV above the barrier the
reduced cross-section for $^{17}$F decreases as compared to $^{19}$F. This decrease is similar to the one 
observed for $^{16}$O as compared to $^{18}$O.
The observed trend can be interpreted as the influence of {\em sd} protons and neutrons on 
the reduced cross-section. In this framework,
at energies upto 20 MeV above the barrier the presence 
of valence {\em sd} neutrons acts to increase the effective size of the system that fuses {\em above the increase in the static size}. Both the $^{16}$O/$^{18}$O and $^{17}$F/$^{19}$F manifest this behavior.
Comparison of a nucleus with 
{\em sd} protons and neutrons to one with {\em sd} neutrons alone indicates 
that the presence of the {\em sd} protons results in a decrease of the effective size. This behavior
is interpreted as the strong interaction between the {\em sd} 
protons and the {\em sd} neutrons. It should be emphasized that this explanation of the trends noted
is qualitative and a quantitative description which requires an accurate description of pairing effects
is not yet available.
Both a widely-used analytic model of fusion and a state-of-the-art 
dynamical model were used to investigate the systematic behavior expected. While both models predict a
decrease in the cross-section with removal of {\em sd} neutrons,
the magnitude of the observed reduction in the fusion cross-section
for $^{17}$F as compared to $^{19}$F is not reproduced. Coupled-channels
calculations which consider only
the ground state nuclei are able to reproduce the measured excitation functions despite the
presence of the low-lying 2s$_{1/2}$ proton-halo state.
This initial observation of the sensitive interplay 
of valence neutrons and protons in the fusion of {\em sd} shell nuclei motivates further 
high-quality measurements of fusion for neutron-rich light nuclei. A new generation of radioactive beam facilities 
\cite{FRIB, GANIL}, 
and in particular the availability of low-energy reaccelerated beams, provides an unprecedented
opportunity to explore this topic and improve our understanding of low-density nuclear matter.

\section{Acknowledgements}

This work was supported by the 
U.S. Department of Energy under Grant Nos. DE-FG02-88ER-40404 (Indiana University), and
the National Science Foundation under Grant No PHY-1491574 (Florida State University).
J.V. acknowledges the support of a NSF Graduate Research Fellowship under Grant No. 1342962.




\bibliographystyle{elsarticle-num}
%\bibliography{F17C12}

%\begin{thebibliography}{00}

%\end{thebibliography}

\end{document}